%% file: 00-main-arXiv.tex
\documentclass[10pt]{article}
\usepackage{tgpagella} 

\usepackage{moresize}
\usepackage{tabularx} 
\usepackage{tablefootnote}
\usepackage[dvips]{graphics}
\usepackage[utf8]{inputenc}
\usepackage{amsmath}
\usepackage{amsfonts}
\usepackage{amssymb,gensymb}
\usepackage{graphicx}
\usepackage{fancyhdr} 
\usepackage{multirow}
\usepackage{wrapfig, framed}
\usepackage{algorithm}
\usepackage{algorithmic}
\usepackage{epsfig,endnotes,color,url,paralist, multirow,float}

\usepackage{subfig}
\usepackage[export]{adjustbox}

\usepackage[left=2cm,right=2cm,top=2cm,bottom=2cm]{geometry}
\usepackage[inline,shortlabels]{enumitem}
\usepackage{mathtools}
\usepackage{mathrsfs} 
\usepackage[table]{xcolor}

\definecolor{lightgray}{gray}{0.9}
\definecolor{LightCyan}{rgb}{0.88,1,1}

\usepackage{lipsum}
\usepackage{comment}
\usepackage{soul}
\usepackage{tikz}
\usepackage{xcolor}
\usepackage[format=plain,
            labelfont={bf,it},
            textfont=it]{caption}
\usepackage{xspace}
\usepackage{hyperref}
\usepackage{booktabs}

\usepackage{cite}
\usepackage{url}

\usepackage[compact]{titlesec}

\makeatletter
\newcommand\mysize{\@setfontsize\notsotiny\@vipt\@viipt}
\makeatother

\newcommand{\ouralg}{{\texttt{ComputeAmp}}\xspace}

\begin{document}

\newpage
\pagenumbering{arabic}
\pagestyle{plain}
\setcounter{page}{1}
\setcounter{section}{0}

\begin{center}{{\bf \Large
Maximizing Compute Capacity in AI Data Centers through Cooling, Energy Storage, and Computing Adaptation}} 
\end{center}


\begin{table}[!h]
\centering
\begin{tabular}{m{0.28\textwidth} m{0.28\textwidth}m{0.28\textwidth}}
\centering Shaolei Ren\\ \emph{UC Riverside}
&\centering Mohammad A. Islam\\ \emph{UT Arlington}
& \centering Adam Wierman\tablefootnote{A preliminary shorter version of this paper is published and presented as a poster at ACM e-Energy 2026. E-mail: Shaolei Ren (shaolei@ucr.edu),
Mohammad A. Islam (mislam@uta.edu),
and Adam Wierman (adamw@caltech.edu)}\\ \emph{Caltech}
\end{tabular}
\end{table}

\begin{center}
    \textbf{Abstract}
    \end{center}

\input{1-abstract}

\vspace{0.3cm}

\input{1-introduction}



{
       \bibliographystyle{unsrt}

\input{reference.bbl}
}

\end{document}

%% file: 1-abstract.tex
The deployment of artificial intelligence is increasingly constrained by limited 
site-level power capacity, which must support both compute systems and non-compute systems (primarily cooling) at all times.
Cooling power demand, especially in non-evaporative cooling systems, can increase substantially with ambient temperature in the summer, producing recurring periods of elevated cooling power that often lasts for multiple hours per day.
Therefore, maximizing compute capacity under a limited site-level power budget is an important planning and operational challenge.
Sizing the compute system conservatively based on peak cooling power can leave part of the site-level power capacity underutilized when the cooling power is below its peak, particularly in cooler months. On the other hand, sizing the compute system aggressively based on low cooling power can cause the total site-level power demand to exceed the site-level power capacity during hot days in the summer.
This paper proposes \ouralg (Compute Amplifier), a framework that maximizes
the compute capacity by jointly and dynamically leveraging cooling, battery energy storage, and computing-based adaptation. 
We discuss the opportunities and limitations of \ouralg and illustrate its potential to significantly expand usable compute capacity within local power and water resource limits. We also present a problem formulation for \ouralg and highlight a few algorithmic and operational challenges.

%% file: 1-introduction.tex
\section{Introduction}\label{sec:introduction}

The rapid growth of artificial intelligence (AI) is fueling the expansion of power-intensive data centers, many of which are being developed at the scale of tens to hundreds of megawatts, with some approaching gigawatt scale \cite{DataCenter_Energy_EPRI_AI_17Percent_US_2030_WhitePaper_2026}.
This has begun to place additional pressure on power grids that are already facing reliability and capacity challenges.
 In many cases, however, the binding constraint for data center grid interconnection is not the total annual electricity demand, but 
the maximum site-level \emph{total} power budget for both compute systems and
non-compute overheads (mostly cooling).
Therefore, maximizing compute capacity under a limited site-level power budget is an important planning and operational challenge.

One challenge is that, under a site-level total power budget, the power available for compute can decrease substantially in summer because elevated ambient temperatures increase cooling power demand, especially in non-evaporative cooling systems without effective mitigation strategies \cite{DataCenter_PeakPower_Water_Tradeoff_AdvancesAppliedEnergy_2026_VANZETTEN2026100269}. The high power density of AI servers often requires liquid cooling, such as direct-to-chip cooling, with closed-loop cooling infrastructure. This server-level cooling loop moves heat from the servers to the facility level, but the heat must still be rejected to the outdoor environment. Without evaporative or adiabatic assistance, outdoor heat rejection relies primarily on sensible heat transfer to ambient air. Thus, during hot periods, higher ambient temperatures can reduce the effectiveness of dry heat rejection and need higher fan power, higher compressor power, or other mechanical cooling support to maintain the required coolant temperature. This leads to elevated cooling power demand in the summer.

\begin{figure}[!h]    \centering
        \subfloat[Richmond, VA]{\includegraphics[width=0.35\textwidth]{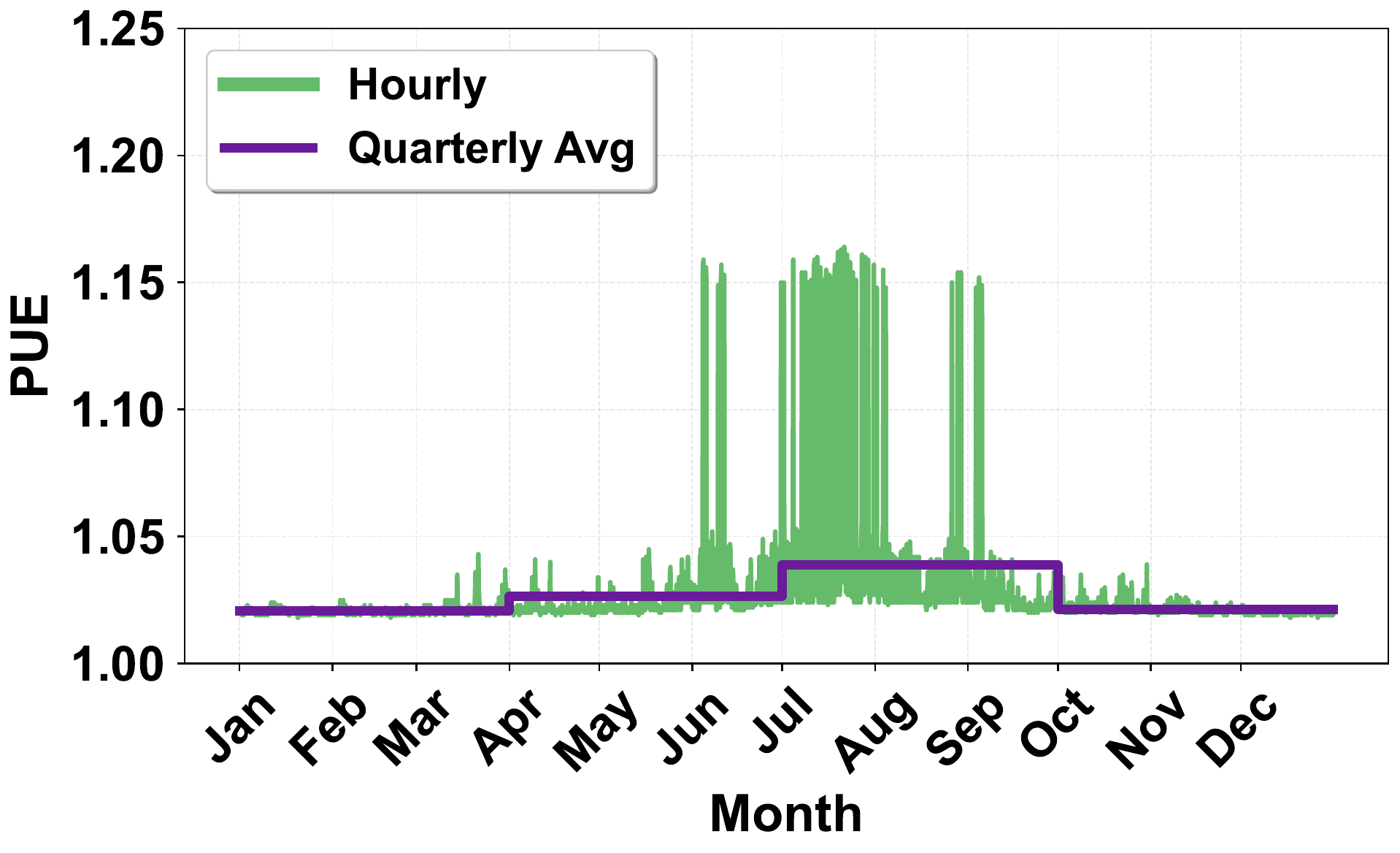}\label{fig:pue_virginia}
        }
        \subfloat[Austin, TX]{
        \includegraphics[width=0.35\textwidth]{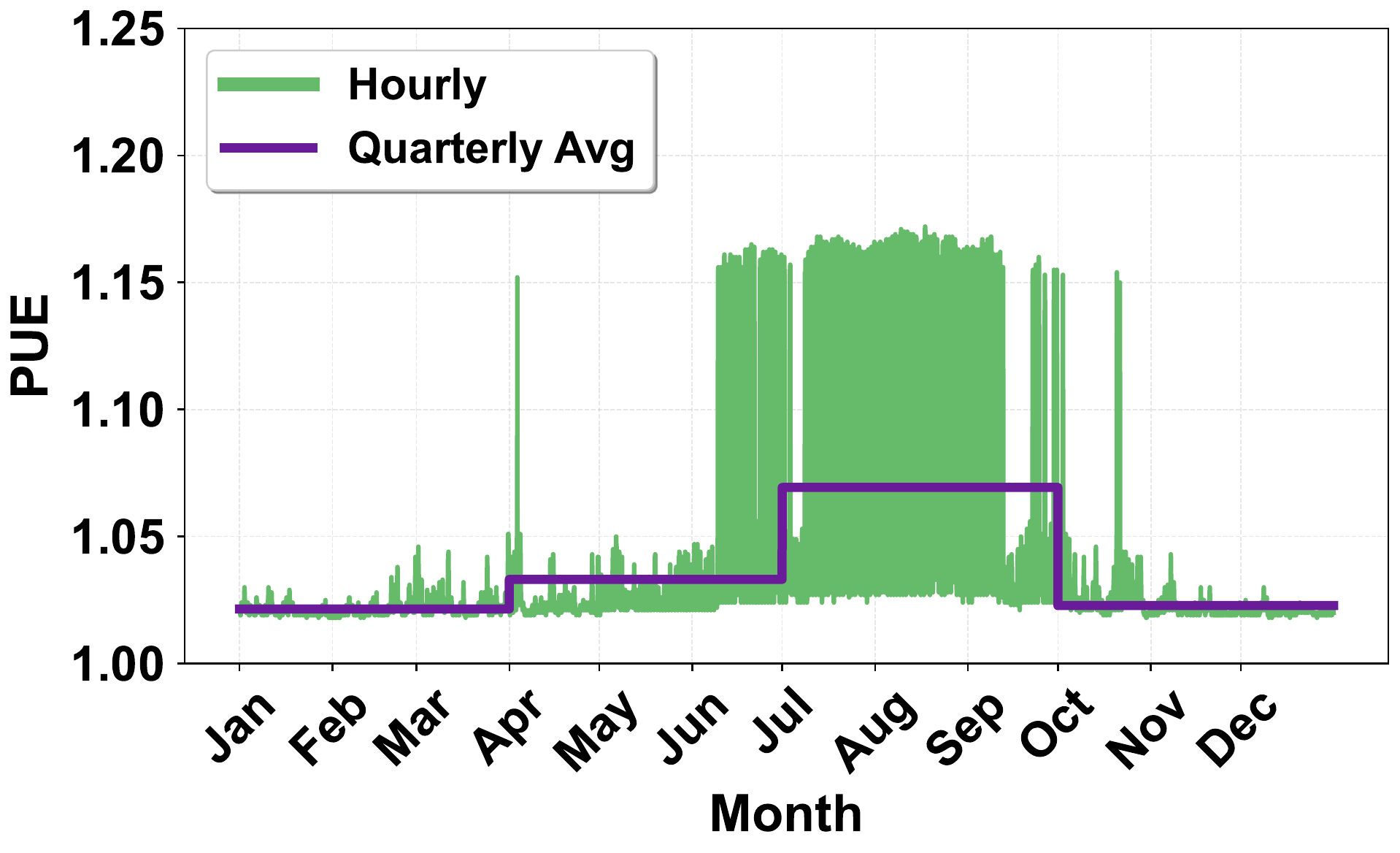}\label{fig:pue_texas}
    }%
        \vspace{-0.1cm}
    \caption{Simulated hourly PUE of a highly-optimized AI data center without adiabatic assistance in Virginia and Texas, respectively, reproduced from~\cite{DataCenter_Cold_UTES_ThermalEnergyStorage_PeakPower_AndrewChien_eEnergy_2026}, indicating substantially higher cooling power in the summer.}\label{fig:hourly_pue}
\end{figure}

Figure~\ref{fig:hourly_pue} shows the simulated hourly power usage effectiveness (PUE, defined as the ratio of total site-level energy to IT energy), for a liquid-cooled AI data center configured according to best practices and without adiabatic assistance.  
While the annualized PUE remains below 1.05, the peak hourly PUE frequently exceeds 1.15 during the hottest hours of the year when cooling efficiency decreases substantially. Crucially, such high-PUE periods are not confined to a few extreme hours each year; rather, they often occur for 5--12 hours per day over multiple hot months, effectively creating seasonal spikes in site-level power demand and reducing the power capacity available for compute.

The seasonal variation in cooling power creates an important challenge for sizing compute capacity within a limited site-level power budget, particularly as large-scale data centers face increasing regional power capacity constraints \cite{DataCenter_Energy_EPRI_AI_17Percent_US_2030_WhitePaper_2026}. 
The challenge is further amplified as the peak cooling power demand typically coincides with periods of high summer grid stress. 

Even if grid-side power capacity constraints are resolved or alleviated in the future, the total data center power capacity will remain a high-value asset, supported by costly energy storage and/or backup generation for high uptime and reliability. 
This is particularly crucial for multi-tenant colocation data centers, which account for more than half of total U.S. data center energy use \cite{DoE_DataCenter_EnergyReport_US_2024}. In such facilities,
 operators do not directly own or control tenant servers,
 and the power infrastructure capital expense often can represent more than half of the total cost of ownership over a 15- to 20-year lifespan \cite{Shaolei_Colocation_Power_Capping_HPCA_2016}.
Therefore, even without grid constraints, addressing seasonal cooling power variation is crucial for efficiently utilizing site-level power capacity for compute.

The challenge posed by seasonal cooling power variation differs from non-firm power arrangements, under which data centers may accept limited load-reduction requirements as a way to obtain faster grid interconnection
\cite{Power_FlexibleDataCenter_NonfirmCapacity_PrincetonZERO_Report_2025_brancucci2025flexible}. Non-firm capacity constraints are typically episodic (e.g., a few hours) and can often be addressed through short-duration measures, such as workload shifting, battery discharge, and in some cases backup generation where permitted. In contrast, cooling-driven power challenges are tied to recurring seasonal weather patterns and may persist for multiple hours per day over extended summer periods. As a result, they are often harder to manage using short-duration flexibility alone and can have a more sustained effect on compute capacity planning, server utilization, and/or workload performance.

One common strategy is to conservatively size compute capacity  based on peak cooling power in the summer, ensuring that the site-level power capacity is not exceeded even during the hottest periods. This approach is operationally straightforward and avoids summer throttling, but it leaves a substantial share of power capacity underutilized during cooler seasons, particularly in winter when the cooling power is much lower. On the other hand, an alternative, albeit not commonly used yet, strategy is to aggressively provision servers based on winter conditions with low cooling power, thereby utilizing more of the available site-level power capacity
for compute. Although this strategy improves site-level power capacity utilization, it may require throttling or shedding computing loads in the summer as cooling power increases, leaving installed server capacity underutilized. As a result, workload performance and service quality can decline during periods of high cooling power.

While purely dry cooling exhibits significant seasonal variation in power use
and adds challenges to effective site-level power capacity utilization, evaporative assistance can help mitigate the peak cooling power demand in the summer.
In practice, evaporative assistance can reduce peak hourly PUE by roughly 10\% \cite{DataCenter_PeakPower_Water_Tradeoff_AdvancesAppliedEnergy_2026_VANZETTEN2026100269} to 30\% \cite{DataCenter_PeakPower_Water_Tradeoff_Model_OregonWashington_Report_2025_pae2025energywater} relative to purely dry cooling. 
For example, a leading technology company reports that, in a Central European climate, evaporative cooling can lower the peak PUE from 1.35 to 1.10 compared with a highly-optimized dry cooling system that needs supplemental mechanical cooling during hot periods \cite{Google_Water_Energy_Tradeoff_Response_to_EU_2026_EuropeanCommission_DataCentreEnergyEfficiencyRating_2026}. 
Such reductions can translate into substantial additional usable compute capacity under a fixed site-level power budget, while easing pressure on grid capacity that could otherwise support broader electrification needs. In a high-growth scenario \cite{DataCenter_Energy_EPRI_AI_17Percent_US_2030_WhitePaper_2026}, the adoption of evaporative assistance by all U.S. data centers constructed after 2023 could yield over 10--30~GW of additional nominal compute capacity by 2030 relative to an all-dry cooling trajectory. This would be sufficient to support roughly 15--45 million additional high-end NVIDIA H100 GPUs. 

Therefore, 
 water can act as a form of power capacity. Under certain operating conditions, limited water use for evaporative assistance can reduce cooling power demand and increase the residual power available for computing under a fixed site-level power budget. Nonetheless, local water infrastructure constraints and other water-use limitations must also be considered, so evaporative assistance should be evaluated as part of a broader power-water tradeoff rather than as a universal solution.

This motivates our proposal of \ouralg (Compute Amplifier), which reduces seasonal cooling power spikes and recovers otherwise stranded compute capacity by jointly leveraging three complementary adaptation mechanisms under local resource constraints:
\begin{enumerate}
    \item \textbf{Cooling:} Evaporative or adiabatic assistance paired with water tanks based on local climate and real-time water stress levels, 
    which can reduce cooling power and thus free more capacity for compute;
    \item \textbf{Battery:} Battery energy discharge, which can shave peak grid draw and provide flexibility over short time scales;
    \item \textbf{Computing:} Workload scheduling, deferral, throttling, or power-performance control, which can align computing demand with dynamic power constraints.
    \end{enumerate}

Unlike existing approaches that often treat water availability, cooling design, and compute adaptation separately, \ouralg takes an integrated and dynamic perspective. 
Conventional cooling and water planning are typically based on static assumptions or worst-case design conditions \cite{DataCenter_PUE_Beyond_PowerScarcity_Verrus_Report_2025_verrus2025beyondpue}, while compute adaptation is often considered in isolation from facility-level cooling and water constraints \cite{Shaolei_GeoDistributedDataCenter_Demand_Response_TSG_2015,DataCenter_Cold_UTES_ThermalEnergyStorage_PeakPower_AndrewChien_eEnergy_2026}. 
Moreover, mitigation strategies for water-related risks have been relatively under-explored.
Water availability is often less dispatchable than electricity generation and can vary significantly with weather, drought conditions, seasonal demand, and local infrastructure constraints. Securing a large firm water supply for data center cooling is subject to physical infrastructure constraints and sometimes permitting challenges, particularly when certain host communities may face water restrictions or infrastructure stress. Over the lifecycle of a data center, increasing climate variability may lead to intermittent droughts and water stress even in historically water-abundant regions, creating temporary challenges for reliable water supply. For example, as of May 19, 2026, drought affected 99.26\% of  Florida, with 22.26\% of the state classified under the most severe category of ``Exceptional Drought'' \cite{US_DroughtMonitor_UNL}.
Therefore, non-firm or conditionally available water supply should be explicitly incorporated into future data center operations, allowing evaporative or adiabatic cooling to be enabled selectively based on time-varying local water availability and constraints. 
 \ouralg leverages real-time operating conditions, including local water-stress information, to dynamically inform cooling, battery and workload decisions, with the goal
of efficiently utilizing site-level power capacity under local power and water resource limits.

\section{The Design of \ouralg}

In this section, we discuss the opportunities and challenges of \ouralg\ by examining 
each adaptation mechanism in \ouralg. We will also present a problem formulation and highlight a few algorithmic and operational challenges for future research.

\subsection{Adaptation Mechanism}

\ouralg addresses the seasonal site-level power variation by leveraging three complementary adaptation mechanisms, as described below.

\subsubsection{Cooling}

The high power density of AI servers often necessitates liquid cooling (e.g., liquid-to-chip) combined with closed-loop cooling technologies. Such server-level cooling, also known as the technology cooling loop, removes heat from the servers to prevent overheating, but that heat must still be rejected to the outdoor environment. At that stage, a tradeoff arises, especially during the hottest periods when ambient air alone becomes less effective for dry heat rejection: refrigerant-based or other non-evaporative heat rejection can avoid direct water use but generally requires more electricity, whereas evaporative or adiabatically assisted cooling consumes water while significantly reducing electricity demand (e.g., by 25--35\% in some climates \cite{Amazon_Water_400m_25_35per_PeakPowerReduction_AI_Cloud_Louisiana_PressRelease_2026,DataCenter_PeakPower_Water_Tradeoff_AdvancesAppliedEnergy_2026_VANZETTEN2026100269,DataCenter_PeakPower_Water_Tradeoff_Model_OregonWashington_Report_2025_pae2025energywater}). Indeed, as noted by one major technology company, water is ``the most efficient means of cooling in many places'' \cite{Google_SustainabilityReport_2025}. 

Other complementary technologies, such as underground thermal energy storage \cite{DataCenter_Cold_UTES_ThermalEnergyStorage_PeakPower_AndrewChien_eEnergy_2026}, can also help smooth cooling power demand, but often involve deployment challenges at scale. In this paper, we focus on commonly-used adiabatic or evaporative assistance, which can reduce cooling power demand during hot periods with relatively low annual water use. Furthermore, onsite water storage tanks may help smooth peak water withdrawals and reduce pressure on local water systems \cite{Shaolei_Water_SmallBottle_BigPipe_arXiv_2026}.

\begin{figure*}[!h]
    \centering
    \subfloat[Quarterly PUE]{
        \includegraphics[height=0.185\textwidth]{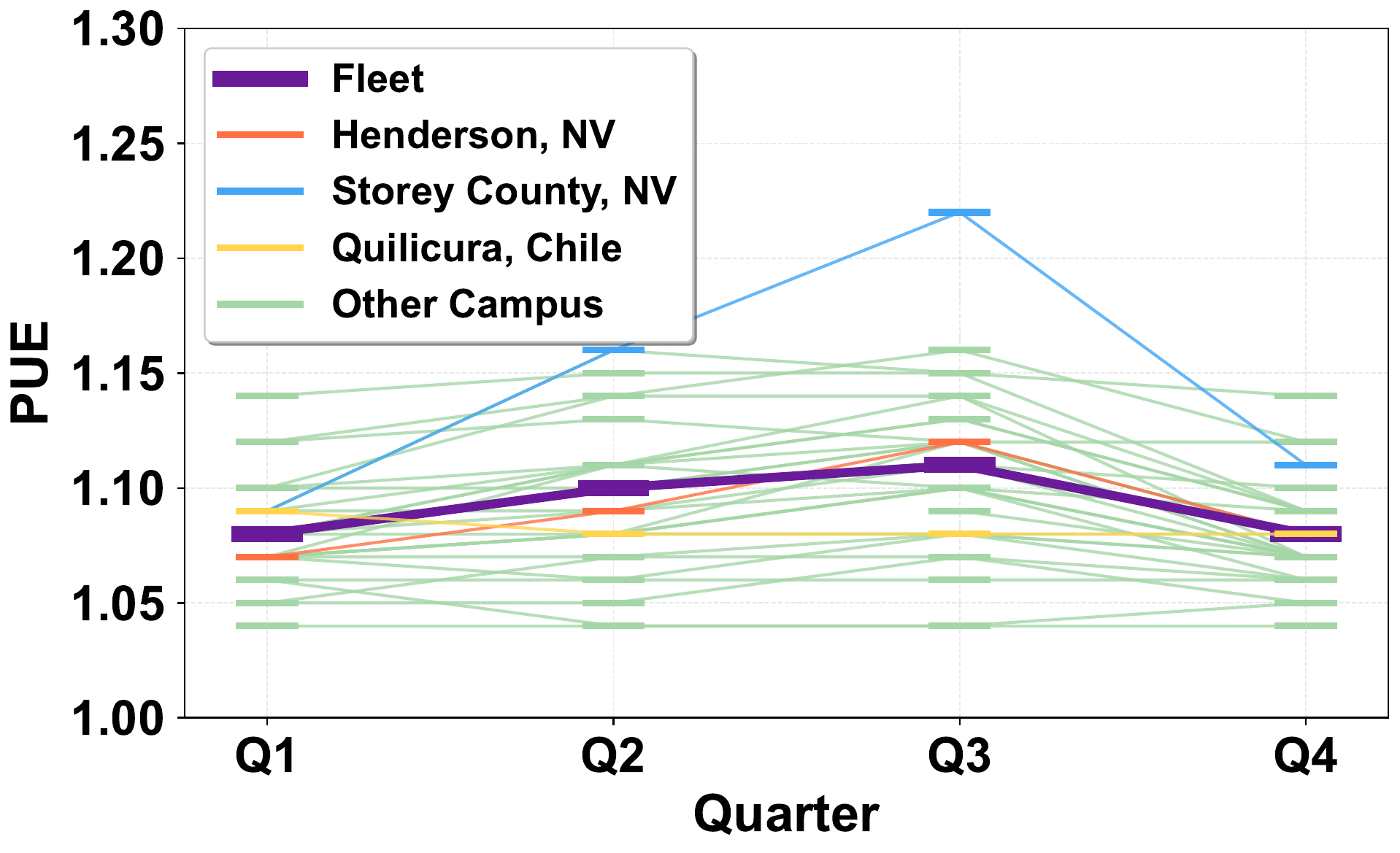}\label{fig:pue_google}
    }%
    \subfloat[Monthly PUE in Arizona]{
        \includegraphics[height=0.185\textwidth]{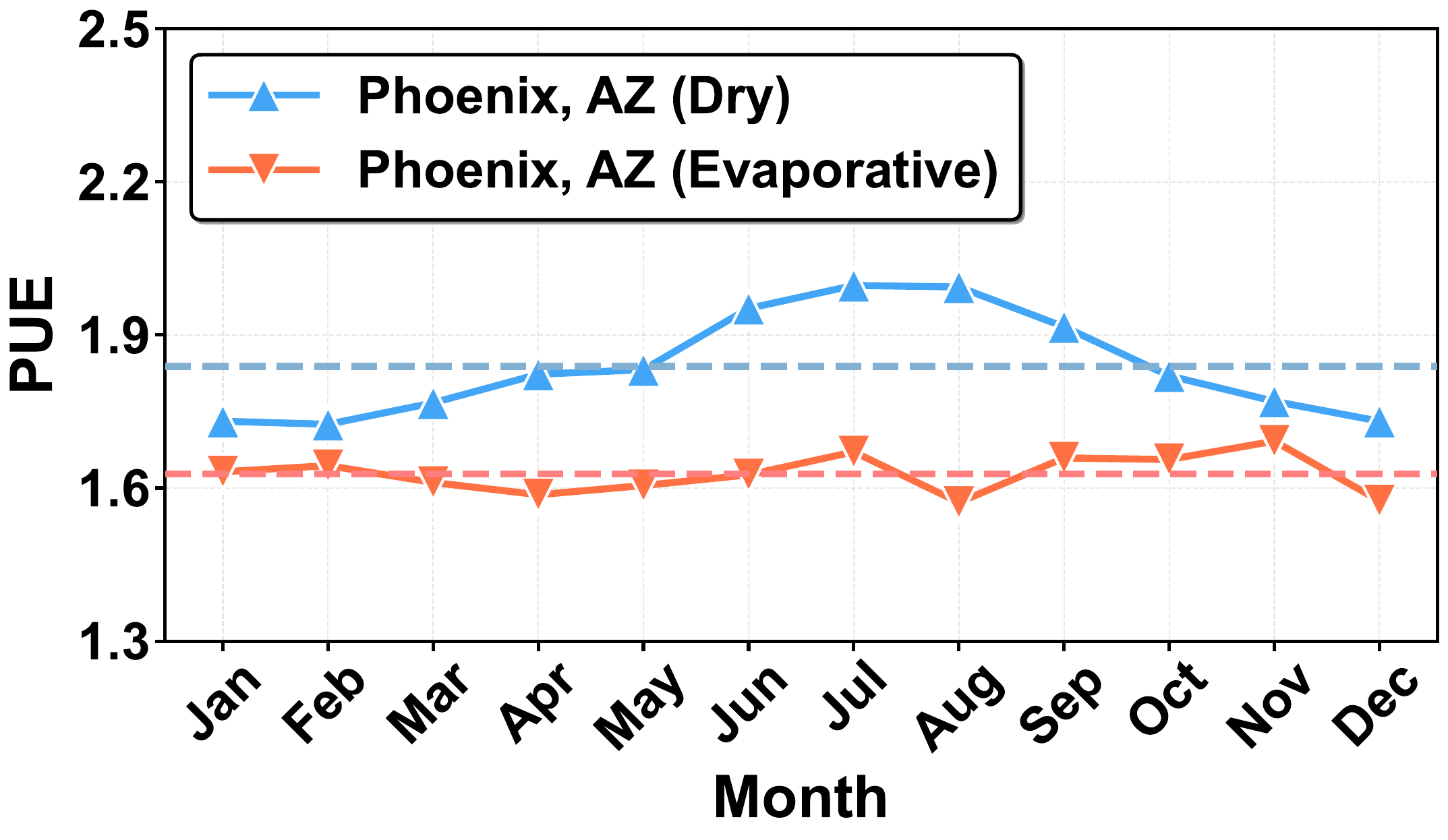}\label{fig:pue_arizona}
    }%
    \subfloat[Hourly PUE]{
        \includegraphics[height=0.185\textwidth]{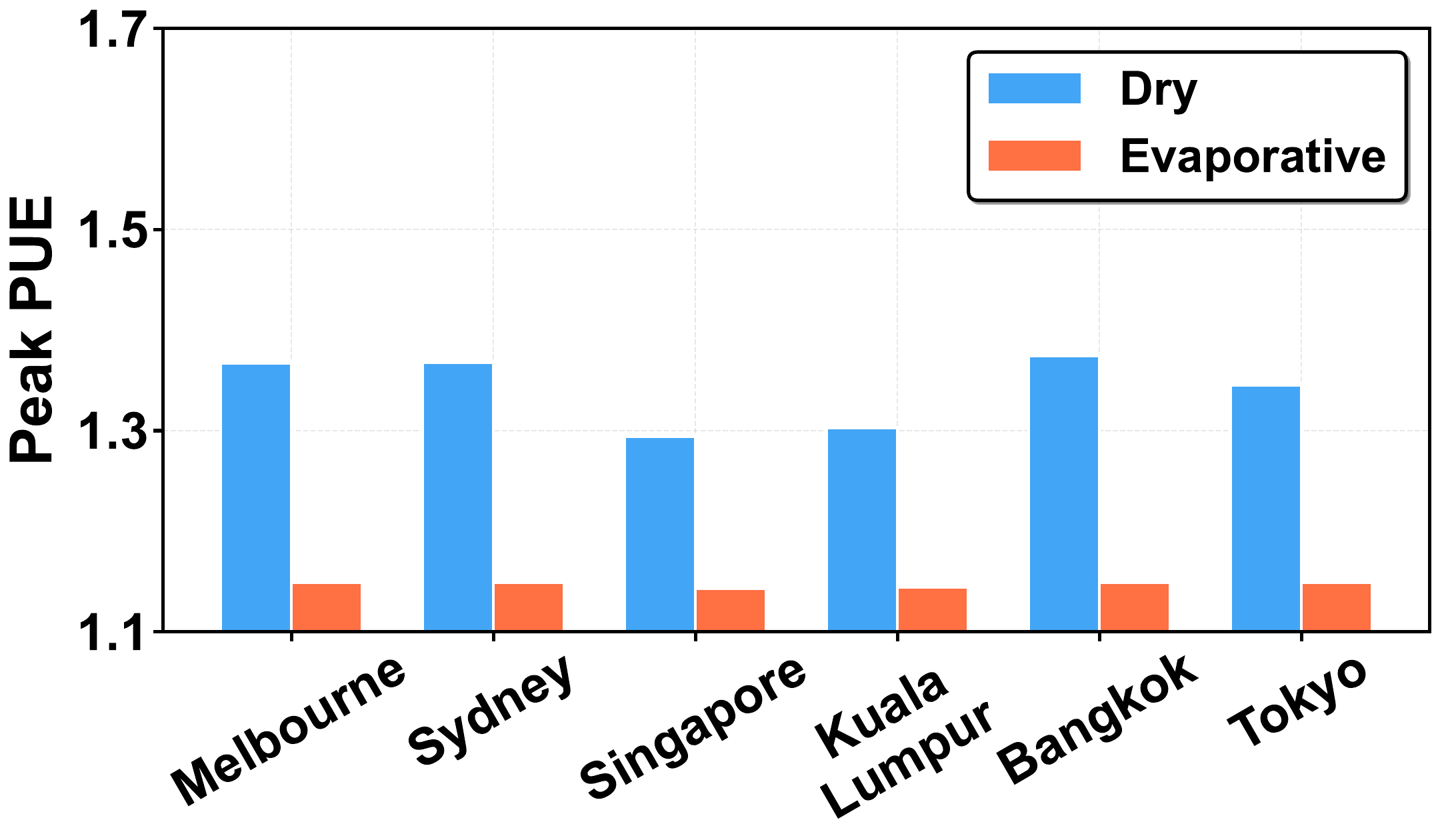}\label{fig:pue_peakd_ltc}
    }%
    \caption{(a) Quarterly PUE of a major technology company in 2025 ~\cite{Google_DataCenter_Efficiency_Website_QuarterlyPUE}. 
    The Storey County, NV, campus uses air-cooled heat rejection, while the other campuses use evaporative or adiabatic assisted heat rejection. 
    The Quilicura, Chile campus is located in 
    the Southern Hemisphere.  
    (b) Monthly PUE in 2019 for two large colocation data centers in Phoenix, AZ with dry cooling and evaporative cooling, respectively \cite{Water_DataCenterEnergy_Tradeoff_Arizona_Real_Measurement_WUE_Monthly_2022_KARIMI2022106194}.
   (c) Simulated hourly peak PUE of a liquid-cooled data center using air-cooled (Dry) or water-cooled chillers (Evaporative) for heat rejection 
~\cite{DataCenter_PeakPower_Water_Tradeoff_AdvancesAppliedEnergy_2026_VANZETTEN2026100269}.}
    \label{fig:pue_comparison}
\end{figure*}

Figure~\ref{fig:pue_comparison} compares the PUE of waterless and evaporative heat rejection. Across quarterly, monthly, and hourly timescales, evaporative cooling can deliver substantial reductions in peak cooling power, thereby enabling greater compute capacity without additional grid capacity expansion. For example, Figure~\ref{fig:pue_google} shows that the air-cooled site in Storey County, NV, exhibits more spikier summer PUE than evaporatively-cooled sites, including Henderson, NV, despite the latter's significantly hotter climate. The company operates several other air-cooled sites  
without evaporative assistance, whose quarterly PUEs are not publicly available.

What matters for data center compute capacity planning is the peak power usage. In practice, evaporative assistance can commonly reduce peak hourly PUE by roughly 10\% \cite{DataCenter_PeakPower_Water_Tradeoff_AdvancesAppliedEnergy_2026_VANZETTEN2026100269} to 30\% \cite{DataCenter_PeakPower_Water_Tradeoff_Model_OregonWashington_Report_2025_pae2025energywater} relative to purely dry cooling, releasing 10--30\% more power capacity for compute. As a result, if evaporative assistance were hypothetically adopted across all U.S. data centers built after 2023, it could unlock more than 10--30~GW of additional nominal compute capacity by 2030 under a high-growth scenario \cite{DataCenter_Energy_EPRI_AI_17Percent_US_2030_WhitePaper_2026}, compared with a fully non-evaporative cooling pathway. 

Nonetheless, for a large data center with several hundred megawatts, evaporatively assisted cooling can use millions of gallons of water per day, often from potable supplies. Depending on the local water infrastructure, this demand may create siting and permitting challenges \cite{Shaolei_Water_SmallBottle_BigPipe_arXiv_2026}. Accordingly, evaporative assistance should be evaluated in relation to local available water capacity and real-time water stress, with non-potable water sources preferred where feasible \cite{EPA_Water_ReuseActionPlan20_Website}.

Evaporative assistance illustrates a partial substitution between water and power under certain operating conditions. This does not imply that water should always be used to reduce electricity consumption, or that electricity should always be used to reduce water use, since water availability, power availability, infrastructure capacity, and local stress conditions vary substantially across locations and time. Rather, it suggests that power and water should be evaluated jointly when seasonal cooling power demand limits the compute capacity available under a fixed site-level power budget.

\subsubsection{Battery}

To ensure high uptime, battery energy storage system (BESS) is a standard component of data centers.  
Depending on its installed capacity, BESS can provide short-term ride-through during power interruptions before on-site generators start, as well as support during demand response events and, in some cases, sustain operations through grid outages lasting up to a few hours. It can also help smooth intermittent power variations caused by large synchronous load swings during AI training \cite{Datacenter_AI_Training_PowerStabilization_Microsoft_Nvidia_OpenAI_arXiv_Oct_2025_choukse2025powerstabilizationaitraining}.
 
 In principle, BESS can also serve as a fast and controllable mechanism for shaving cooling-driven power spikes in the summer by discharging during hot periods and recharging at night when cooling demand is lower.
 In practice, however, the limited number of cooler nighttime hours in hot regions such as Texas and Nevada, together with charging-rate constraints, can render BESS alone insufficient to offset prolonged cooling-driven peak power.
Moreover, installing a new dedicated large BESS for seasonal cooling power shaving is often costly \cite{DataCenter_Cold_UTES_ThermalEnergyStorage_PeakPower_AndrewChien_eEnergy_2026}, while repurposing the existing BESS originally reserved for site backup can undermine the desired reliability targets because seasonal cooling power spikes are recurrent, prolonged, and high in magnitude.
Therefore, BESS is often better suited to shave the \emph{residual} cooling-power spikes that remain after applying evaporative assistance, where applicable.

\subsubsection{Computing}

Computing-based flexibility, including workload shifting and server speed throttling, has been extensively studied as a means of handling 
power emergencies in oversubscribed data centers and maximizing the utilization of a given \emph{fixed} compute power capacity \cite{Fan:2007:PPW:1250662.1250665}. 
More recently, it has also emerged as a potential mechanism for short-term demand response under occasional grid stress, which typically occurs for a few hours per year. Nonetheless, such computing-based adaptation often incurs performance degradation due to server slowdown, workload deferral, and related operational disruptions \cite{AdamWierman_DataCenterDemandResponse_Survey_IGCC_2014}.
It is particularly challenging for multi-tenant colocation data centers, which represent a key segment of the industry but lack a central entity with control over all the servers for coordinated adaptation \cite{Shaolei_Colo_TruthDR_Infocom_2015}.

While computing-based adaptation may be effective for addressing short-term power constraints, 
the performance penalties can make it  
less suitable as a standalone solution for addressing prolonged compute power shortfalls caused by seasonal cooling power spikes. Instead, it may be better positioned as a complementary adaptation mechanism alongside
 evaporative assistance and BESS, where applicable.

\subsection{Problem Formulation}

We consider a discrete-time horizon indexed by time slots $t \in \mathcal{T} := \{1,\dots,T\}$, where each slot has a fixed duration (e.g., 15 minutes or 1 hour). 
Thus, for simplicity of presentation, we use power and energy interchangeably without explicitly accounting for the duration of each time slot.
This abstraction assumes that sub-slot power variations are managed by short-duration energy buffers or by separate grid-side balancing mechanisms.

Let $P^{\max}_t$ denote the facility-level power capacity, $p_t^{\mathrm{IT}}$ the IT power consumed by computing servers, $p_t^{\mathrm{cool}}$ the cooling power, $p_t^{\mathrm{BESS}}$ the BESS power, and
$p_t^{\mathrm{grid}}$ the grid power draw,
respectively. We retain the subscript $t$ in the facility-level power capacity $P^{\max}_t$ to allow for the general case in which data centers may be required to reduce power consumption under non-firm power agreements during periods of grid stress; otherwise, $P^{\max}_t$ is equal to the allocated power capacity. The BESS power $p_t^{\mathrm{BESS}}$ is positive when charging and negative when discharging, subject to the rate constraint $p_t^{\mathrm{BESS}}\in[-\bar{P}^{\mathrm{dis}}, \bar{P}^{\mathrm{ch}}]$
where $\bar{P}^{\mathrm{dis}}$ and $\bar{P}^{\mathrm{ch}}$ are the discharging and
charging power limits, respectively.

The site-level total power $p_t^{\mathrm{grid}}$ drawn from the grid is
\begin{equation}
p_t^{\mathrm{grid}}
=
p_t^{\mathrm{IT}} + p_t^{\mathrm{cool}} + p_t^{\mathrm{BESS}},
\qquad \forall t \in \mathcal{T},
\label{eq:power_balance}
\end{equation}
subject to the power capacity constraint 
\begin{equation}
p_t^{\mathrm{grid}} \le P_t^{\max}, \qquad \forall t \in \mathcal{T}.
\label{eq:facility_cap}
\end{equation}

\subsubsection{Cooling and water model}

Let $w_t\in[0,W]$ denote the water used for evaporative or adiabatic assistance at time $t$ with $W$ being the largest water flow rate, and let $s_t^{\mathrm{w}}$ denote the state of the water tank. 
Here, for simplicity of presentation, we consider water withdrawal without  accounting for water discharge. 
The cooling power depends on both the cooling load and the amount of evaporative assistance. A generic reduced-form model is
\begin{equation}
p_t^{\mathrm{cool}} = f_t(p_t^{\mathrm{IT}}, w_t),
\label{eq:cooling_model}
\end{equation}
where $f_t(\cdot)$ may depend on ambient weather conditions. For a given $p_t^{\mathrm{IT}}$, $f_t$ is non-increasing in $w_t$, reflecting that greater evaporative assistance can reduce more cooling power up to physical limits. Similarly, for a given $w_t$, $f_t$ is non-decreasing in $p_t^{\mathrm{IT}}$, reflecting that a higher IT load generates more heat and therefore requires more cooling power.

The water tank dynamics are
\begin{equation}
\label{eqn:water_tank_dynamics}
s_{t+1}^{\mathrm{w}} = s_t^{\mathrm{w}} + r_t - w_t, \text{ and  } S_{\min}^{\mathrm{w}}\leq s_{t+1}^{\mathrm{w}} \leq S_{\max}^{\mathrm{w}} \qquad \forall t \in \mathcal{T},
\end{equation}
where $r_t$ is the amount of replenished water during slot $t$ subject to the peak water supply rate, and $S_{\min}^{\mathrm{w}}$ and $S_{\max}^{\mathrm{w}}$
represent the minimum and maximum water tank levels, respectively.

\subsubsection{BESS model}

Let $s_t^{\mathrm{b}}$ denote the BESS state of charge. The BESS dynamics are governed by 
\begin{equation}
s_{t+1}^{\mathrm{b}}
=
s_t^{\mathrm{b}}
+
\eta^{\mathrm{ch}} \left[p_t^{\mathrm{BESS}}\right]^+ 
-
\frac{1}{\eta^{\mathrm{dis}}}
\left[-p_t^{\mathrm{BESS}}\right]^+, \text{ and }
{S}_{\min}^{\mathrm{b}} \leq s_t^{\mathrm{b}} \leq {S}_{\max}^{\mathrm{b}}
\qquad \forall t \in \mathcal{T},
\label{eq:battery_dyn}
\end{equation}
where $[x]^+ = \max\{x,0\}$, and
$\eta^{\mathrm{ch}}, \eta^{\mathrm{dis}} \in (0,1]$ are the charging and
discharging efficiencies, 
and ${S}_{\min}^{\mathrm{b}}$ and ${S}_{\max}^{\mathrm{b}}$
are the minimum and maximum BESS states of charge, respectively.

The BESS model is  simplified for the convenience of presentation, while detailed BESS models can be incorporated by making the charging and discharging efficiencies, degradation cost, or feasible power range depend on the state of charge \cite{Battery_Impact_ChargingRate_EV_2021_mothilal2021impact}.

\subsubsection{Workload model}

We model the IT power demand directly rather than explicitly representing workload arrivals, queues, or workload-shifting dynamics. Let $p_t^*$ denote the baseline IT power demand in slot $t$ without power adaptation. The realized IT power $p_t^{\mathrm{IT}}$ is constrained by
\begin{equation}
\underline{p}_t^{\mathrm{IT}} \le p_t^{\mathrm{IT}} \le p_t^*,
\qquad \forall t \in \mathcal{T},
\label{eq:it_power_constraints}
\end{equation}
where $\underline{p}_t^{\mathrm{IT}}$ represents the minimum IT power needed to satisfy service-quality requirements in slot $t$. This abstraction captures the aggregate effect of workload and server-side adaptations, such as workload shifting, throttling, power capping, and admission control, without explicitly modeling workload dynamics. In this paper, we focus on the interaction between IT power demand, cooling power, and BESS operation, leaving detailed workload-level modeling to future work.

\subsubsection{Optimization Objective}

While minimizing energy cost is important, the primary goal of \ouralg is to coordinate cooling, BESS, and IT power adaptation to satisfy site-level power constraints while limiting water use, BESS operation, and IT power reduction. 
Thus, we focus on the following weighted total cost minimization objective, while allowing other cost considerations to be incorporated where applicable: 
\begin{equation}
\min \sum_{t=1}^{T}
\left[
\lambda_t^{\mathrm{w}} w_t
+ \lambda_t^{\mathrm{b}} \left|p_t^{\mathrm{BESS}}\right|
+ \lambda_t^{\mathrm{p}} C_t^{\mathrm{perf}}\bigl(p_t^* - p_t^{\mathrm{IT}}\bigr)
\right],
\label{eq:objective}
\end{equation}
where $\lambda_t^{\mathrm{w}}\geq0$ is a time-varying weight on water use that captures temporal variation in water-stress indicators, $w_t$ is the amount of water used for evaporative assistance, $\lambda_t^{\mathrm{b}}\geq0$ captures the BESS operating or degradation cost per unit of energy throughput, 
 $C_t^{\mathrm{perf}}(\cdot)\geq0$ represents the non-decreasing workload performance penalty associated with reducing IT power below its baseline demand $p_t^*$ and
 $\lambda_t^{\mathrm{p}}\geq0$ is the weight for performance penalty. The penalty function $C_t^{\mathrm{perf}}\bigl(p_t^* - p_t^{\mathrm{IT}}\bigr)$ can capture the aggregate effect of workload shifting, throttling, power capping, admission control, or other workload- and server-side adaptation mechanisms.

Put together, the problem solved by \ouralg over the horizon $\mathcal{T}$ can be written as:
\begin{align}
\min_{\{p_t^{\mathrm{IT}},\, w_t,\, p_t^{\mathrm{BESS}}\}_{t \in \mathcal{T}}}
\quad &
\sum_{t=1}^{T}
\left[
\lambda_t^{\mathrm{w}} w_t
+ \lambda_t^{\mathrm{b}} \left|p_t^{\mathrm{BESS}}\right| + \lambda_t^{\mathrm{p}} C_t^{\mathrm{perf}}\bigl(p_t^* - p_t^{\mathrm{IT}}\bigr)
\right]
\label{eq:full_problem}
\\
\text{s.t.} \quad
&
p_t^{\mathrm{IT}} + f_t(p_t^{\mathrm{IT}}, w_t) + p_t^{\mathrm{BESS}}
\le P_t^{\max},
\qquad \forall t \in \mathcal{T},
\label{eq:cons_power_capacity_compact}
\\
&
\underline{p}_t^{\mathrm{IT}} \le p_t^{\mathrm{IT}} \le p_t^*,
\qquad \forall t \in \mathcal{T},
\label{eq:cons_it_power_compact}
\\
&
\text{Water tank dynamics in \eqref{eqn:water_tank_dynamics}},
\label{eq:cons_water_compact}
\\
&
\text{BESS dynamics in \eqref{eq:battery_dyn}}.
\label{eq:cons_battery_compact}
\end{align}

The formulation is general and can be parameterized based on site-specific and/or time-varying constraints. For example, if a location faces temporary or persistent water scarcity, or other constraints on water use, $\lambda_t^{\mathrm{w}}$ can be set to a sufficiently large value so that evaporative assistance is effectively disabled during the affected periods. 
Similarly, if IT power reduction is not feasible, such as in a multi-tenant colocation data center where the operator may not directly control server power, the performance-penalty weight $\lambda_t^{\mathrm{p}}$ can be set to a sufficiently large value to discourage or effectively disable IT power adaptation. In this case, the formulation relies primarily on evaporative assistance and BESS operation to satisfy the site-level power constraints.

The formulation can also be extended to a multi-site setting by incorporating geographical load balancing across data centers \cite{Shaolei_Water_SpatioTemporal_GLB_TCC_2018_7420641}. In this case, workload shifting decisions can be coordinated with site-specific power capacity, cooling conditions, water availability, and BESS states to reduce overall cost while satisfying service and infrastructure constraints.

Although we present the formulation as an operational problem over a horizon $\mathcal{T}$, it can also support planning decisions. In this case, the operational problem in \eqref{eq:full_problem}--\eqref{eq:cons_battery_compact} can be used as a feasibility constraint, while the outer planning problem minimizes planning-level costs such as peak water withdrawal, onsite water-storage capacity, BESS capacity, or the required site-level power budget under representative weather, workload, and water-stress scenarios.

\subsection{Illustrative Results}

To illustrate the tradeoff between BESS and water use, we simulate a highly-optimized AI data center located in Austin, Texas, over the period from July 1 to September 30, 2023.
This period corresponds to the same year considered in \cite{DataCenter_Cold_UTES_ThermalEnergyStorage_PeakPower_AndrewChien_eEnergy_2026} and covers some hottest months of the year.
The results should be interpreted as a \emph{directional illustration}, rather than as a deployment-ready engineering estimate.

\subsubsection{Settings}
The data center follows a workload trace similar to that considered in \cite{DataCenter_Cold_UTES_ThermalEnergyStorage_PeakPower_AndrewChien_eEnergy_2026}, with an average IT load of 84.5~MW and a peak IT load of 98.6~MW.
We do not model compute-side adaptation, as it can be challenging to implement in many multi-tenant colocation data centers where the facility operator typically has no or very limited control over tenants' servers and workloads.

For this illustrative analysis, we do not model grid-side load-curtailment events and instead consider three fixed site-level power capacity constraints:
$P^{\max}=100~\mathrm{MW}$ (low),
 $P^{\max}=105~\mathrm{MW}$ (medium),
 and 
$P^{\max}=110~\mathrm{MW}$ (high). 
We also consider two BESS configurations: a small BESS with an energy capacity of 25~MWh and a large BESS with an energy capacity of 200~MWh. Relative to the peak IT load of 98.6~MW, these capacities provide slightly more than 15 minutes and about two hours of support, respectively. The small-BESS case reflects the current practice of using batteries primarily as an emergency power bridge during power outages, before backup generators are activated. In contrast, the large-BESS case captures the emerging use of batteries for longer-duration demand response.

We consider an idealized cooling model in which all servers are liquid-cooled and heat is rejected through dry coolers. When the ambient dry-bulb temperature is no higher than $35^\circ\mathrm{C}$, the dry coolers are assumed to operate with an effective PUE no greater than 1.05. The air flow rate is sized based on a $10^\circ\mathrm{C}$ air-side temperature rise. When the ambient dry-bulb temperature exceeds $35^\circ\mathrm{C}$, we consider two alternatives. The first is dry cooling with supplemental mechanical refrigeration, represented by an effective PUE of 1.18 \cite{DataCenter_Cold_UTES_ThermalEnergyStorage_PeakPower_AndrewChien_eEnergy_2026}. The second is adiabatic assistance, which pre-cools the ambient air toward $35^\circ\mathrm{C}$ before heat rejection through the dry coolers.
For the adiabatic-assisted case, we assume an idealized evaporative pre-cooling process in which water is used only to reduce the inlet air temperature required by the dry coolers. The water requirement is computed from the air flow rate and outdoor psychrometric conditions, including dry-bulb temperature, relative humidity, and wet-bulb temperature. This assumption represents a lower-bound estimate of water use because it neglects non-idealities such as incomplete evaporation, drift, blowdown, distribution losses, and control inefficiencies. In addition, real systems often require lower temperature setpoints than the assumed $35^\circ\mathrm{C}$ here, especially for air-cooled servers or AI hardware operated under tighter thermal-performance constraints (e.g., for higher performance via safe overclocking) \cite{Microsoft_Cooling_FlexibleCool_AI_TemperatureLower_Blog_April_2026}. Similarly, the assumed peak PUE of 1.18 for the mechanically assisted dry-cooling case is optimistic for many real deployments.
For example, a leading technology company reports that, in a Central European climate, even a highly optimized mechanically cooled system has a peak PUE of 1.35 \cite{Google_Water_Energy_Tradeoff_Response_to_EU_2026_EuropeanCommission_DataCentreEnergyEfficiencyRating_2026}.
Therefore, the results should be interpreted as a \emph{directional illustration} of the BESS--water tradeoff rather than as a site-specific engineering estimate.

\begin{figure}[!t]    \centering
        \subfloat[Water demand vs. battery SOC (25 MWh)]{\includegraphics[width=0.33\textwidth]{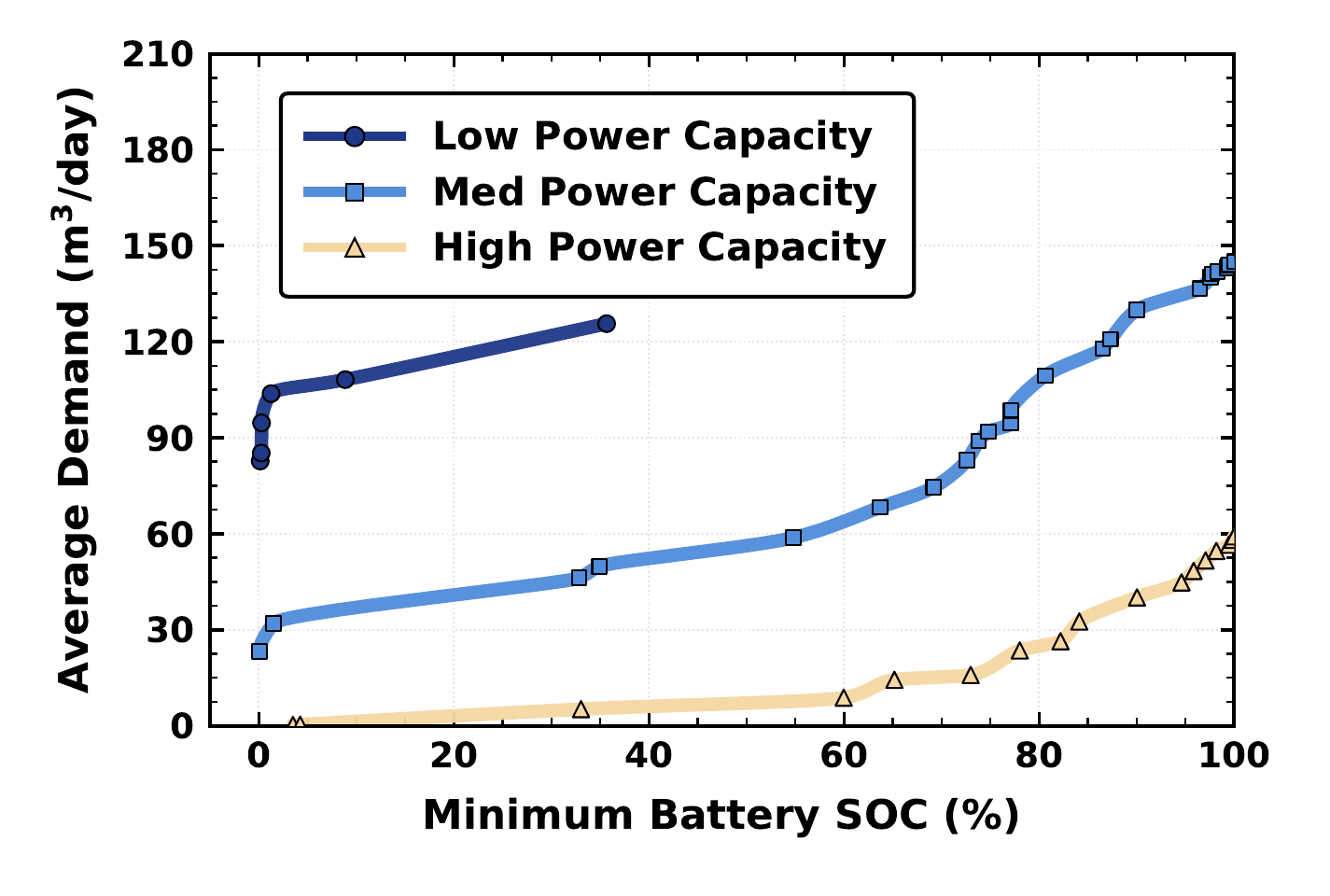}\label{fig:water_battery_tradeoff_25mwh}
        }
        \subfloat[Water demand vs. battery SOC (200~MWh)]{\includegraphics[width=0.33\textwidth]{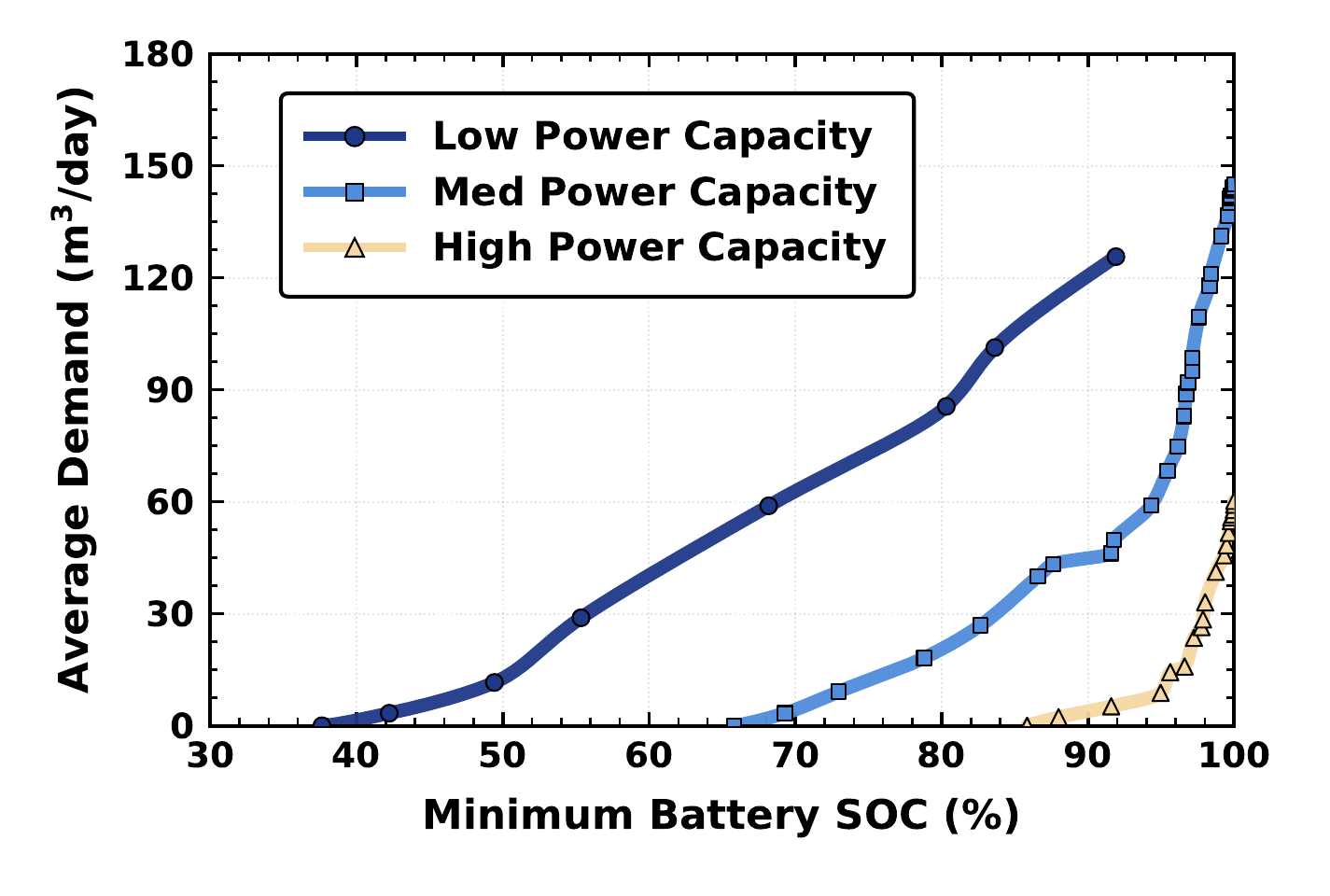}\label{fig:water_battery_tradeoff_200mwh}
        }
        \subfloat[Peak water demand vs. water tank size]{
        \includegraphics[width=0.33\textwidth]{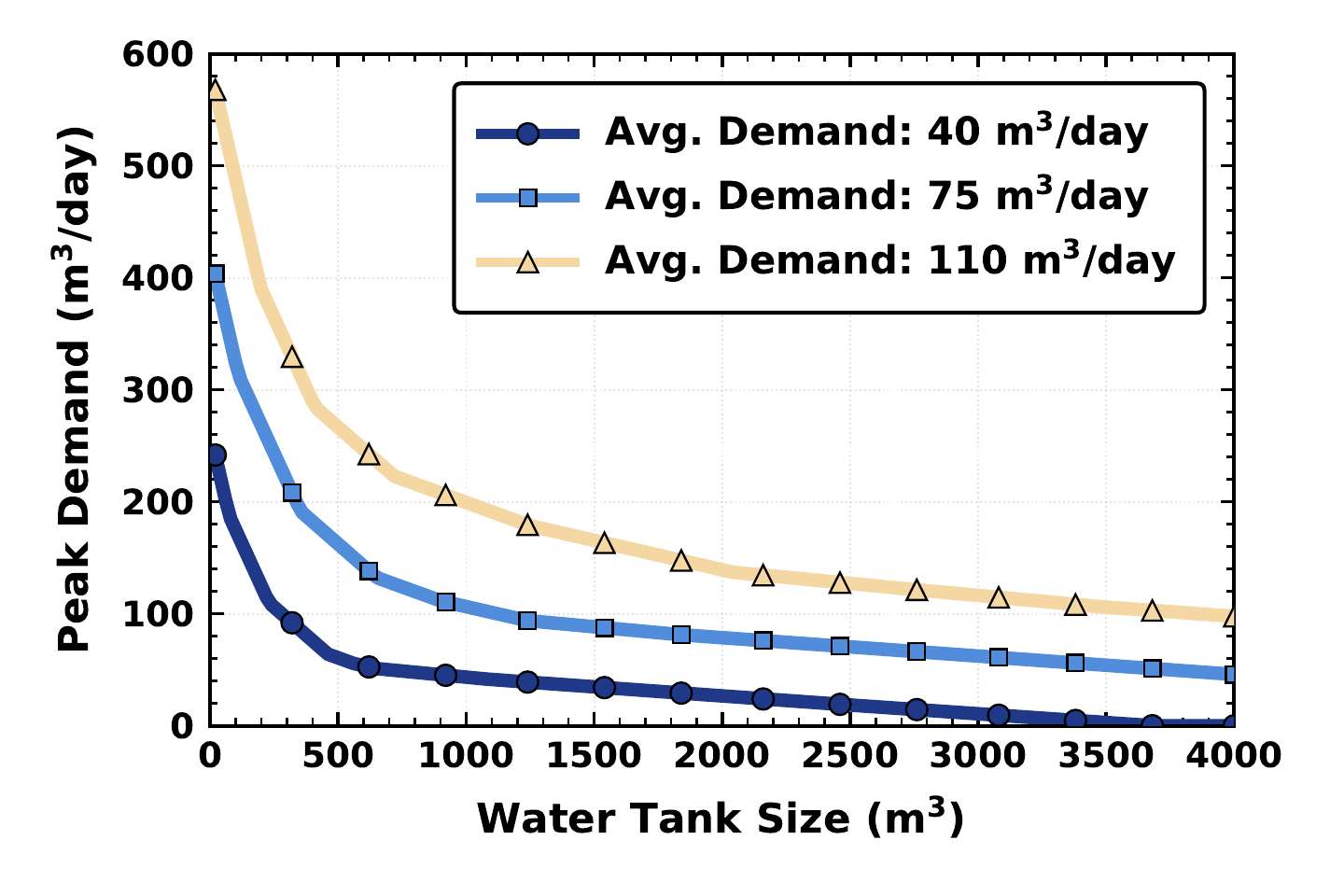}\label{fig:water_tank_size}
    }%
        \vspace{-0.1cm}
        \caption{Illustrative results for an AI data center simulated under idealized settings in Austin, TX, from July 1 to September 30, 2023.}\label{fig:illustrative}
\end{figure}

\subsubsection{Water and BESS tradeoff}
Figures~\ref{fig:water_battery_tradeoff_25mwh} and~\ref{fig:water_battery_tradeoff_200mwh} show the relationship between water use and BESS dispatch under different site-level power-capacity constraints for the small-BESS and large-BESS settings, respectively. In our evaluation, adiabatic assistance and BESS discharge serve as two complementary mechanisms for satisfying the site-level power constraint. Together, they create a tradeoff between water consumption and BESS use, where BESS use is reflected by the minimum state of charge (SOC). A lower minimum SOC indicates deeper battery discharge and, therefore, greater potential site reliability risk.

Under a tight site-level power-capacity constraint, adiabatic assistance alone may not reduce power usage enough to keep the site within its capacity limit, making BESS discharge necessary. Conversely, when the BESS is small and the site-level power capacity is low, BESS discharge alone may be insufficient because the battery can be depleted during high-demand periods. In such cases, adiabatic assistance becomes necessary to further reduce cooling power and prevent the total site load from exceeding the capacity limit. Thus, the operation must carefully balance BESS energy discharge and water use. Deeper BESS discharge can help satisfy the site-level power constraint but may reduce backup energy availability, whereas greater use of adiabatic assistance can lower cooling power at the cost of additional water consumption.

Because the analysis adopts idealized assumptions, the maximum amount of BESS energy usage or water required is relatively small. As the site-level power capacity constraint becomes more stringent, the required reliance on water and/or BESS increases, moving the tradeoff curve outward.

\subsubsection{Peak water demand and water tank size tradeoff}
Next, we consider the medium site-level power-capacity case, $P^{\max}=105~\mathrm{MW}$, under three different average water use levels. We examine how water storage can smooth peak water demand and thereby reduce potential pressure on local water infrastructure, particularly during hot periods when other users may also experience elevated water demand. The results are shown in Figure~\ref{fig:water_tank_size}. 
The water tank size is optimized so that cumulative water use never exceeds the sum of cumulative water supply and the stored water level. This represents an idealized scenario. In practice, additional storage headroom is often needed to account for forecast uncertainty, operational constraints, and resilience considerations. 

In general, increasing the water tank size substantially reduces peak water demand from the local water system. When the average water use level is low, a sufficiently large water tank can further reduce or even eliminate the need for incremental water supply during the periods of adiabatic assistance, assuming the tank can be pre-filled when system-wide water demand is lower. Although these results are based on idealized assumptions, they suggest that water storage can be a cost-effective mechanism for smoothing data centers' peak water demand and reducing their short-term impacts on local community water systems.

\subsection{Algorithmic and Operational Challenges}

The formulation of \ouralg in \eqref{eq:full_problem}--\eqref{eq:cons_battery_compact} 
relies on future information that may not be fully known at the time decisions are made. For example, cooling power depends on future ambient weather conditions, water use decisions may depend on time-varying water-stress indicators, and IT power adaptation depends on future workload demand and service requirements. In practice, these quantities are typically obtained through forecasts rather than known exactly. However, forecast errors can lead to either insufficient adaptation, causing violations of site-level power constraints, or overly conservative decisions, reducing compute utilization or increasing unnecessary water and BESS use.

This creates a need for algorithms that can combine prediction with robustness gaurantees. One possible direction is learning-augmented optimization \cite{Shaolei_Learning_SOCO_Delay_NeurIPS_2023,OnlineOpt_Learning_Augmented_RobustnessConsistency_NIPS_2020,L2O_LearningMLAugmented_Regression_CR_GeRong_NIPS_2021_anand2021a}, in which forecasts of weather, workload demand, water stress, and grid constraints are used to improve empirical performance, while the algorithm is designed to retain some robustness when predictions are inaccurate.
For example, the algorithm may use predicted high-temperature periods to schedule evaporative assistance or reserve BESS capacity in advance, while maintaining safety margins to handle forecast errors. Similarly, workload adaptation can be planned using predicted demand, but with online correction as actual workload and cooling conditions are observed.

Another challenge is model uncertainty. For example, the workload performance penalty model and the cooling power model may not be perfectly known in advance and may vary over time as workload mix, server operating conditions, cooling equipment behavior, and ambient conditions change. These models therefore need to be estimated and updated online using operational data. This leads to a joint learning and control problem, where the system must make operational decisions while improving its estimates of workload performance impacts and cooling-power responses. Reinforcement learning provides one possible framework for this setting \cite{Reinforcement_Book_2018_sutton2018reinforcement_intro}, although safety constraints, limited exploration, and the need to satisfy site-level power at all times are research challenges
for deployment.

Operational implementation depends on the flexibility available at a given site. The formulation does not require all adaptation mechanisms to be available; instead, unavailable or undesirable mechanisms can be discouraged or effectively disabled through the corresponding penalty weights or constraints. For example, evaporative assistance may be dynamically enabled, disabled, or modulated based on local water conditions and cooling needs, while IT power adaptation may be limited or disabled in settings such as multi-tenant colocation data centers where the operator does not directly control server power. BESS operation must then be coordinated with the available mechanisms, for example by reserving battery energy for periods when water use is limited and/or residual cooling power spikes remain high. These considerations suggest that \ouralg is not only an optimization formulation, but also a control and systems-integration solution involving forecasting, online decision-making, and site-specific infrastructure capabilities.

\section{Conclusion}

\ouralg\ provides a framework for increasing usable compute capacity in power-constrained AI data centers by addressing seasonal variation in cooling power demand. Under a fixed site-level power capacity, elevated cooling power demand during hot periods can reduce the power available for IT loads, while provisioning for peak cooling demand can leave power capacity underutilized during cooler periods. \ouralg\ coordinates three complementary adaptation mechanisms: evaporative assistance to reduce cooling power demand, BESS operation to mitigate residual power constraints, and IT power adaptation to reduce compute load when needed.
By setting appropriate parameters, \ouralg can be tuned to reflect site-specific conditions, such as temporary water-use limitations or limited workload flexibility, and can also support planning decisions involving peak water demand, BESS capacity, and site-level power budgets. While our model is simplified, it provides a basis for studying how cooling, BESS, and workload-side flexibility can be coordinated to improve compute utilization within infrastructure constraints.

More broadly, \ouralg suggests a research direction for AI data center operation and planning: treating usable compute capacity as a function of cooling, water, BESS, and workload decisions. In this view, water and power can become partially substitutable resources in some operating regimes, where limited water use for evaporative assistance can reduce cooling power demand and increase the power available for computing. This highlights the need to explicitly navigate power-water tradeoffs rather than optimizing either resource in isolation.

Future work is needed to develop robust and learning-augmented algorithms, improve cooling and workload performance models using operational data, and extend \ouralg to multi-site settings through geographical load shifting.